\documentstyle[preprint,aps,epsfig]{revtex}
\tighten
\begin{document}
\draft
%__________________________________________________
%\twocolumn
%_______________________ Title, Authors ____________________________________
\preprint{\parbox[t]{80mm}{Preprint Number: 
        \parbox[t]{45mm}{ANL-PHY-8920-TH-98
%                                             \\ nucl-th/ddmmnnn
}}}

\title{Pseudovector components of the pion, $\pi^0\to \gamma\gamma$, and
$F_\pi(q^2)$}

\author{Pieter Maris and Craig D. Roberts} 
\address{Physics Division, Bldg. 203, Argonne National Laboratory,
Argonne IL 60439-4843}
\date{\today}
\maketitle
%-------------------------------------------------------------------
\begin{abstract}
As a consequence of dynamical chiral symmetry breaking the pion
Bethe-Salpeter amplitude necessarily contains terms proportional to
$\gamma_5\,\gamma\cdot P$ and \mbox{$\gamma_5\,\gamma\cdot k$}, where $k$ is
the relative and $P$ the total momentum of the constituents.  These terms are
essential for the preservation of low energy theorems, such as the
Gell-Mann--Oakes-Renner relation and those describing anomalous decays of the
pion, and to obtaining an electromagnetic pion form factor that falls as
$1/q^2$ for large $q^2$, up to calculable $\ln q^2$-corrections.  In a simple
model, which correlates low- and high-energy pion observables, we find $q^2
F_\pi(q^2) \sim 0.12$ - $0.19\,$GeV$^2$ for
$q^2 \mathrel{\rlap{\lower4pt\hbox{\hskip0pt$\sim$}}
\raise2pt\hbox{$>$}} 10\,{\rm GeV}^2$.
\end{abstract}

% 11.10.St,       Bound and unstable states, BSEs
% 12.38.Lg,       Other non.pert. calculations
% 12.39.Pn,       Potential models
% 12.40.Yx,       Hadron mass models and calculations
% 13.40.Gp        Electromagnetic form factors
% 14.40.-n,       Mesons
% 14.40.Aq,       pi K and eta mesons
% 24.85.+p        Quarks, gluons, and QCD in nuclei and nuclear processes 
%..............................

%\twocolumn

% \section{Introduction}
% \hspace*{-\parindent}{\bf Introduction.}\hspace*{\parindent} 
%
\section{Pion as a bound state}
Understanding the pion is a key problem in strong interaction physics.  As
the lowest mass excitation in the strong interaction spectrum it must provide
the long-range attraction in \mbox{$N$-$N$} potentials~\cite{bob95}.  In QCD,
it is a quark-antiquark bound state whose low- and high-energy properties
should be understandable in terms of its internal structure, and it is also
that nearly-massless, collective excitation which is the realisation of the
Goldstone mode associated with dynamical chiral symmetry breaking (DCSB).  An
explanation of these properties requires a melding of the study of the many
body aspects of the QCD vacuum with the analysis of two body bound states.
The Dyson-Schwinger equations (DSEs) provide a single, Poincar\'e invariant
framework that is well suited to this problem.

The DSEs are a system of coupled integral equations and truncations are
employed to define a tractable problem.  In truncating the system it is
straightforward to preserve the global symmetries of a gauge field
theory~\cite{brs96} and, although preserving the local symmetry is more
difficult, progress is being made~\cite{ayse97}.  The approach has been
applied extensively~\cite{dserev} to the study of confinement, and to DCSB
where the similarity between the ground state of QCD and that of a
superconductor can be exploited, with the QCD gap equation realised as the
quark DSE.  It has also been employed in studying meson-meson and
meson-photon interactions~\cite{pctrev}, heavy meson decays~\cite{ivanov},
QCD at finite temperature and density~\cite{thermo}, and strong interaction
contributions to weak interaction phenomena~\cite{ewff,hecht}.

Studying the pion as a bound state requires an understanding of its
(fully-amputated) Bethe-Salpeter amplitude, which has the general form
\begin{eqnarray}
\label{genpibsa}
\Gamma_\pi^j(k;P) & = & \tau^j \gamma_5 \left[ i E_\pi(k;P) + \gamma\cdot P
F_\pi(k;P) \rule{0mm}{5mm}\right. \\ \nonumber & & \left. \rule{0mm}{5mm}+
\gamma\cdot k \,k \cdot P\, G_\pi(k;P) + \sigma_{\mu\nu}\,k_\mu P_\nu
\,H_\pi(k;P) \right]\,,
\end{eqnarray}
where $\{\tau^j\}_{j=1\ldots 3}$ are the Pauli matrices.  $\Gamma_\pi^j$
satisfies the renormalised, homogeneous Bethe-Salpeter equation
\begin{eqnarray}
\label{genbse}
\left[\Gamma_\pi^j(k;P)\right]_{tu} &= & 
\int^\Lambda_q  \,
[\chi_\pi^j(q;P)]_{sr} \,K^{rs}_{tu}(q,k;P)\,,
\end{eqnarray}
where $k$ is the relative and $P$ the total momentum of the quark-antiquark
pair, $\chi_\pi^j(q;P) := S(q_+) \Gamma_\pi^j(q;P) S(q_-)$, $r$,\ldots,$u$
represent colour, flavour and Dirac indices, $q_\pm=q\pm P/2$, and
$\int^\Lambda_q := \int^\Lambda d^4 q/(2\pi)^4$ represents mnemonically a
translationally-invariant regularisation of the integral, with $\Lambda$ the
regularisation mass-scale: $\Lambda \to \infty$ is the last step in any
calculation.  In Eq.~(\ref{genbse}), $K$ is the fully-amputated, renormalised
quark-antiquark scattering kernel and $S$ is the renormalised dressed-quark
propagator, which is the solution of
\begin{eqnarray}
\label{gendse}
S(p)^{-1} & = & Z_2 (i\gamma\cdot p + m_{\rm bm})
+\, Z_1 \int^\Lambda_q \,
g^2 D_{\mu\nu}(p-q) \gamma_\mu S(q)
\Gamma_\nu(q,p) \,,
\end{eqnarray}
where $D_{\mu\nu}(k)$ is the renormalised dressed-gluon propagator,
$\Gamma_\mu(q;p)$ is the renormalised dressed-quark-gluon vertex and $m_{\rm
bm}(\Lambda)$ is the Lagrangian current-quark bare mass.  In
Eq.~(\ref{gendse}), $Z_1$ and $Z_2$ are the renormalisation constants for the
quark-gluon vertex and quark wave function, and the chiral limit is obtained
with $m_{\rm bm}(\Lambda)=0$.  The solution of Eq.~(\ref{gendse}) has the
general form
\begin{eqnarray}
\label{sp}
S(p) & = & -i\gamma\cdot p\, \sigma_V(p^2) + \sigma_S(p^2) 
         \equiv  \frac{1}{i \gamma\cdot p\,A(p^2) + B(p^2)}\,.
\end{eqnarray}

Also important in the study of the pion is the chiral-limit, axial-vector
Ward-Takahashi identity
\begin{equation}
\label{avwti}
-i P_\mu \Gamma_{5\mu}^{j}(q;P) = 
S^{-1}(q_+) \,\gamma_5 \frac{\tau^j }{2} +
         \gamma_5 \frac{\tau^j}{2} \,S^{-1}(q_-) \,.
\end{equation}
This identity relates the renormalised, dressed-quark propagator to the
renormalised axial-vector vertex, which satisfies
\begin{eqnarray}
\label{genave}
\left[\Gamma_{5\mu}^j(k;P)\right]_{tu} & = &
Z_2 \, \left[\gamma_5\gamma_\mu \frac{\tau^j}{2}\right]_{tu} \,+
\int^\Lambda_q \, [\chi_{5\mu}^j(q;P)]_{sr} \,K^{rs}_{tu}(q,k;P)\,,
\end{eqnarray}
where $\chi_{5\mu}^j(q;P) := S(q_+) \Gamma_{5\mu}^j(q;P) S(q_-)$.  In the
chiral limit the general solution of Eq.~(\ref{genave}) is~\cite{mrt98,mr97}
\begin{eqnarray}
\label{truavv}
\Gamma_{5 \mu}^j(k;P) & = &
\frac{\tau^j}{2} \gamma_5 
\left[ \rule{0mm}{5mm}\gamma_\mu F_R(k;P) + \gamma\cdot k k_\mu G_R(k;P) 
- \sigma_{\mu\nu} \,k_\nu\, H_R(k;P) \right]\\
&+ & \nonumber
 \tilde\Gamma_{5\mu}^{j}(k;P) 
+\,f_\pi^0\,  \frac{P_\mu}{P^2 } \,\Gamma_\pi(k;P)\,,
\end{eqnarray}
where: $F_R$, $G_R$, $H_R$ and $\tilde\Gamma_{5\mu}^{j}$ are regular as
$P^2\to 0$; $P_\mu \tilde\Gamma_{5\mu}^{j}(k;P) \sim {\rm O }(P^2)$;
$\Gamma_\pi(k;P)$ is the amplitude in Eq.~(\ref{genpibsa}); and the residue
of the pole in the axial-vector vertex is $f_\pi^0$, the chiral-limit
leptonic decay constant, which is obtained from:
\begin{eqnarray}
\label{caint}
\delta^{ij} f_\pi P_\mu = 
Z_2\int^\Lambda_q\,
{\rm tr}\left[\frac{\tau^i}{2} \gamma_5 \gamma_\mu 
S(q_+) \Gamma_\pi^j(q;P) S(q_-)\right]\,.
\end{eqnarray}
[This expression is valid for arbitrary values of the quark mass.]  Now,
independent of assumptions about the the form of $K$, it follows~\cite{mrt98}
from Eqs.~(\ref{sp}), (\ref{avwti}) and (\ref{truavv}) that
\begin{eqnarray}
\label{bwti} 
f_\pi^0 E_\pi(k;0)  &= &  B_0(k^2)\,, \\
\label{awti} 
 F_R(k;0) +  2 \, f_\pi^0 F_\pi(k;0)                 & = & A_0(k^2)\,, \\
G_R(k;0) +  2 \,f_\pi^0 G_\pi(k;0)    & = & 2 A_0^\prime(k^2)\,,\\
\label{gwti} 
H_R(k;0) +  2 \,f_\pi^0 H_\pi(k;0)    & = & 0\,,
\end{eqnarray}
where $A_0$ and $B_0$ are the chiral limit solutions of Eq.~(\ref{gendse}).
A necessary consequence of Eqs.~(\ref{bwti})-(\ref{gwti}) is that the
pseudovector components $F_\pi$ and $G_\pi$, and the pseudotensor component
$H_\pi$, are nonzero in Eq.~(\ref{genpibsa}).  This corrects a
misapprehension~\cite{ds79} that only $E_\pi\neq 0$ and has important
phenomenological consequences.

% \section{Normalisation of the pion field}
% \vspace*{0.5\baselineskip}
% 
% \hspace*{-\parindent}{\bf Normalisation of the pion
% field.}\hspace*{\parindent}
%
\subsection{Normalisation of the pion field}
To highlight one such consequence we note that Eq.~(\ref{genbse}), the
homogeneous Bethe-Salpeter equation, does not determine the normalisation of
the Bethe-Salpeter amplitude.  The canonical normalisation is fixed by
requiring that the pion pole in the quark-antiquark scattering amplitude: $M
:= K / [1 - (SS) K]$, have unit residue.  As an alternative, one can
normalise the solution of Eq.~(\ref{genbse}) by requiring that $E(0;0) =
B(0)$ in the chiral limit.  In terms of the amplitude ${\cal G}_\pi(k;P)$
defined in this way, the canonical normalisation condition is
\begin{eqnarray}
\label{pinorm}
\lefteqn{ 2 \delta^{ij} N_\pi^2\,P_\mu = 
\int^\Lambda_q \left\{\rule{0mm}{5mm}
{\rm tr} \left[ 
\bar{\cal G}_\pi^i(q;-P) \frac{\partial S(q_+)}{\!\!\!\!\!\!\partial P_\mu} 
{\cal G}_\pi^j(q;P) S(q_-) \right]
\right. 
} \\
& & \nonumber \left.  
\;\;\;\;\;\;\;\;\;\;\;\;\;\;\;\;\;\;\;
 + {\rm tr} \left[ 
\bar{\cal G}_\pi^i(q;-P) S(q_+) {\cal G}_\pi^j(q;P) 
        \frac{\partial S(q_-)}{\!\!\!\!\!\!\partial P_\mu}\right]
\rule{0mm}{5mm}\right\}  \\
& & \nonumber
\;\;\;\;\;\;\;\;\;\;
 + \int^\Lambda_{q}\int^\Lambda_{k} \,[\bar\chi_{{\cal G}_\pi}^i(q;-P)]_{sr} 
\frac{\partial K^{rs}_{tu}(q,k;P)}
{\!\!\!\!\!\!\!\!\!\!\!\!\partial P_\mu}\, 
[\chi_{{\cal G}_\pi}^j(k;P)]_{ut}\,,
\end{eqnarray}
where $\bar{\cal G}_\pi(q;-P)^t := C^{-1}\,{\cal G}_\pi(-q;-P)\,C$ with
$C=\gamma_2\gamma_4$, the charge conjugation matrix, and $X^t$ denoting the
matrix transpose of $X$.  Equation~(\ref{pinorm}) defines the pion
normalisation constant, $N_\pi$, which has mass-dimension one.  Physical
observables are expressed in terms of
$\Gamma_\pi(k;P) := (1/N_\pi)\,{\cal G}_\pi(k;P)$.

In the chiral limit, when all the amplitudes in Eq.~(\ref{genpibsa}) are
retained, one obtains~\cite{mrt98}
\begin{equation}
\label{fpinpi}
f_\pi^0 = N_\pi^0\,.
\end{equation}
This result verifies a core assumption in chiral perturbation theory; i.e.,
that the pion field is normalised by $f_\pi^0$, which is implicit in the
expression of the chiral field as
\begin{equation}
U(x):= {\rm e}^{i \vec{\tau}\cdot
\vec{\pi}(x)/f_\pi^0}\,.
\end{equation}
However, Eq.~(\ref{fpinpi}) is violated in bound state
treatments of the pion that neglect the pseudovector components of the
Bethe-Salpeter amplitude~\cite{mr97}.\footnote{$N_\pi$ in
Eq.~(\protect\ref{pinorm}) provides the best numerical approximation to the
pion's leptonic decay constant in analyses that neglect the pseudovector
components and employ a $P$-independent form for $K$.}

% \section{Anomalous pion decay}
% \vspace*{0.5\baselineskip}
% 
% \hspace*{-\parindent}{\bf {\boldmath $\pi^{\mbox{\boldmath $0$}} \to
% \gamma\gamma$} decay.}\hspace*{\parindent}
%
\section{Anomalous Neutral Pion Decay}
The pseudovector components of the pion also play a special role in the
anomalous $\pi^0\to \gamma\gamma$ decay.  Consider the renormalised, impulse
approximation to the axial-vector--photon-photon (AVV) amplitude:\footnote{In
our Euclidean metric: $\gamma_\mu^\dagger = \gamma_\mu$,
$\{\gamma_\mu,\gamma_\nu\}=2 \,\delta_{\mu\nu}$, and a spacelike vector,
$k_\mu$, has $k^2>0$.}
\begin{eqnarray}
\label{calT}
{\cal T}_{\rho\mu\nu}(k_1,k_2) & := & T_{\rho\mu\nu}(k_1,k_2) + 
                                T_{\rho\nu\mu}(k_2,k_1)\,,\\
\label{Trho}
T_{\rho\mu\nu}(k_1,k_2) & := & 
N_c\int^\Lambda_q 
\left\{{\rm tr}_{DF}\left[\rule{0mm}{\baselineskip}
S(q_1)\,\Gamma_{5\rho}^{3}(\hat q;-P)\,
S(q_2)\,i\Gamma^\gamma_\mu(q_2,q_{12})
                S(q_{12})\,i\Gamma^\gamma_\nu(q_{12},q_1)
\right]\right\}\,,
\end{eqnarray}
where $k_1$, $k_2$ are the photon momenta [$k_1^2=0=k_2^2$, $2 k_1\cdot
k_2=P^2$], $q$ is the loop-momentum, and $q_1:= q-k_1$, $q_2:= q+k_2$, $\hat
q:= \case{1}{2}(q_1+q_2)$, $q_{12}:= q-k_1+k_2$.  

Here $\Gamma^\gamma_\mu(p_1,p_2)$ is the renormalised, dressed-quark-photon
vertex, and it is because this vertex satisfies the vector Ward-Takahashi
identity:
\begin{equation}
\label{vwti}
(p_1 - p_2)_\mu \, i\Gamma^\gamma_\mu(p_1,p_2) = 
S^{-1}(p_1) - S^{-1}(p_2)\,,
\end{equation}
that no renormalisation constants appear explicitly in Eq.~(\ref{Trho}).
$\Gamma^\gamma_\mu(p_1,p_2)$ has been much studied~\cite{ayse97} and,
although its exact form remains unknown, its robust qualitative features have
been elucidated so that a phenomenologically efficacious Ansatz has
emerged~\cite{bc80}
\begin{eqnarray}
\label{bcvtx}
i\Gamma^\gamma_\mu(p,q) &:= &
i\Sigma_A(p^2,q^2)\,\gamma_\mu
+ (p+q)_\mu\,\left[\case{1}{2}i\gamma\cdot (p+q) \, \Delta_A(p^2,q^2)
        + \Delta_B(p^2,q^2)\right]\,;\\
\Sigma_f(p^2,q^2) &:= & \case{1}{2}\,[f(p^2)+f(q^2)]\,,\\
\Delta_f(p^2,q^2) & := & \frac{f(p^2)-f(q^2)}{p^2-q^2}\,,
\end{eqnarray}
where $f= A, B$.  A feature of Eq.~(\ref{bcvtx}) is that the vertex is
completely determined by the renormalised dressed-quark propagator.  In
Landau gauge the quantitative effect of modifications, such as that canvassed
in Ref.~\cite{cp92}, is small and can be compensated for by small changes in
the parameters that characterise a given model study~\cite{hawes}.

In the chiral limit ($P^2=0$) using Eqs.~(\ref{genpibsa}) and (\ref{truavv}),
the divergence of the AVV vertex is
\begin{eqnarray}
\label{sepa}
P_\rho\,T_{\rho\mu\nu}(k_1,k_2) & = &
R^3_{\mu\nu}(k_1,k_2) + f_\pi^0 \, T^3_{\mu\nu}(k_1,k_2) \,,
\end{eqnarray}
where the direct contribution from the axial-vector vertex is 
\begin{eqnarray}
\lefteqn{R^3_{\mu\nu}(k_1,k_2) := 
- N_c\int^\Lambda_q 
{\rm tr}_{DF}\left[\rule{0mm}{\baselineskip} S(q_1)\,
\gamma_5\frac{\tau^3}{2}
\left(\rule{0mm}{\baselineskip}
\gamma\cdot P \,F_R(\hat q;0) 
\right.
\right.
}\\
& & \nonumber \left.\left.
+ \gamma\cdot \hat q  \,\hat q\cdot P \,G_R(\hat q;0)
+ \sigma_{\mu\nu}\,\hat q_\mu P_\nu \,H_R(\hat q;0) 
\rule{0mm}{\baselineskip}\right)
S(q_2)\,i\Gamma^\gamma_\mu(q_2,q_{12})
                S(q_{12})\,i\Gamma^\gamma_\nu(q_{12},q_1)
\rule{0mm}{\baselineskip}\right]\,,
\end{eqnarray}
and that from the pion bound state is
\begin{eqnarray}
\lefteqn{T^3_{\mu\nu}(k_1,k_2) := 
 N_c\int^\Lambda_q 
{\rm tr}_{DF}\left[\rule{0mm}{\baselineskip} S(q_1)\,
\gamma_5\tau^3
\left(\rule{0mm}{\baselineskip}
iE_\pi(\hat q; 0) - 
\gamma\cdot P \,F_\pi(\hat q;0) 
\right.
\right.
}\\
& & \nonumber \left.\left.
- \gamma\cdot \hat q  \,\hat q\cdot P \,G_\pi(\hat q;0)
- \sigma_{\mu\nu}\,\hat q_\mu P_\nu \,H_\pi(\hat q;0) 
\rule{0mm}{\baselineskip}\right)
S(q_2)\,i\Gamma^\gamma_\mu(q_2,q_{12})
                S(q_{12})\,i\Gamma^\gamma_\nu(q_{12},q_1)
\rule{0mm}{\baselineskip}\right]\,.
\end{eqnarray}

Using Eqs.~(\ref{awti})-(\ref{gwti}), Eq.~(\ref{sepa}) simplifies:
\begin{eqnarray}
\label{sepb}
\lefteqn{P_\rho\,T_{\rho\mu\nu}(k_1,k_2)  = 
\hat R^3_{\mu\nu}(k_1,k_2) + f_\pi^0 \, \hat T^3_{\mu\nu}(k_1,k_2) \,;}\\
\label{hatR}
\lefteqn{\hat R^3_{\mu\nu}(k_1,k_2) := - N_c\int^\Lambda_q 
{\rm tr}_{DF}\left[\rule{0mm}{\baselineskip} S(q_1)\,
\gamma_5\frac{\tau^3}{2}
\left(\rule{0mm}{\baselineskip}
\gamma\cdot P \,A_0(\hat q^2) 
\right.
\right.}\\
&& \nonumber
\left.
\left.
 + \gamma\cdot \hat q  \,\hat q\cdot P \,2 A_0^\prime(\hat q^2)\right)
S(q_2)\,i\Gamma^\gamma_\mu(q_2,q_{12})
                S(q_{12})\,i\Gamma^\gamma_\nu(q_{12},q_1)
\rule{0mm}{\baselineskip}\right]\,,\\
\label{hatT}
\hat T^3_{\mu\nu}(k_1,k_2) & := &
 N_c\int^\Lambda_q 
{\rm tr}_{DF}\left[\rule{0mm}{\baselineskip} 
S(q_1)\,\gamma_5\tau^3 \, iE_\pi(\hat q; 0) \,
S(q_2)\,i\Gamma^\gamma_\mu(q_2,q_{12})
                S(q_{12})\,i\Gamma^\gamma_\nu(q_{12},q_1) \right]\,.
\end{eqnarray}
Now using Eqs.~(\ref{sp}) and (\ref{bcvtx}) in Eq.~(\ref{hatR}) yields
\begin{equation}
\label{hatRval}
\hat R^3_{\mu\nu}(k_1,k_2) =\,-\,
\frac{\alpha_{\rm em}}{\pi}\,
\epsilon_{\mu\nu\rho\sigma} k_{1\rho} k_{2\sigma}\, {\cal R}(P^2=0)
\end{equation}
where ($s:=q^2$, $A_0^\prime:= \case{d}{ds}A_0(s)$, etc.),
\begin{eqnarray}
{\cal R}(0) & = & \int_0^\infty\,ds\,s^2\,
A_0^2\,\sigma_V^0\,\left(
A_0\left[ (\sigma_V^{0})^2
        + s\,\sigma_V^{0} \,\sigma_V^{0\,\prime} + 
                \sigma_S^{0\,\prime}\,\sigma_S^{0} \right]
+ \sigma_V^{0}\left[s A_0^\prime\,\sigma_V^{0} 
        + B_0^\prime\,\sigma_S^{0}\right] \right)\\
\label{hatRzero} & \equiv & 0\,.
\end{eqnarray}
The last line follows because, using Eq.~(\ref{sp}) to eliminate $\sigma_V^0$
and $\sigma_S^0$ in favour of $A_0$ and $B_0$, the integrand is identically
zero.  Hence the pseudovector components of the neutral-pion Bethe-Salpeter
amplitude {\em combine} with the regular pieces of the axial-vector vertex to
generate that part of the AVV vertex which is {\em conserved}.

To reveal the anomalous contribution to the divergence, consider
Eq.~(\ref{hatT}), in which using Eqs.~(\ref{sp}) and (\ref{bcvtx}) yields
\begin{eqnarray}
\label{hatTval}
\lefteqn{\hat T^3_{\mu\nu}(k_1,k_2)  = 
\frac{\alpha_{\rm em}}{\pi}\,
\epsilon_{\mu\nu\rho\sigma} k_{1\rho} k_{2\sigma}\, {\cal T}(0)\,,}\\
\label{calTO}
{\cal T}(0) & = &  
\int_0^\infty\,ds\,s\,E_\pi\,A_0\,\sigma_V^0\,
\left( A_0\left[ \sigma_V^0\,\sigma_S^0 + 
                s\left(\sigma_V^{0\,\prime}\,\sigma_S^0
                        - \sigma_V^0\,\sigma_S^{0\,\prime}\right)
        + s\,\sigma_V^0\,\left(A_0^\prime \,\sigma_S^0
                                - B_0^\prime \,\sigma_V^0\right) \right]
\right)\,.
\end{eqnarray}

Now, introducing $C(s):= B_0(s)^2/[s A_0(s)^2]$, Eq.~(\ref{calTO}) simplifies
\begin{equation}
{\cal T}(P^2=0) = 
-\int_0^\infty\,ds\,s\,\frac{E_\pi(s;0)}{B_0(s)}\,
        \frac{C^\prime(s)}{[ 1 + C(s)]^3}\,,
\end{equation}
which, using Eq.~(\ref{bwti}), yields
\begin{equation}
f_\pi^0 \,{\cal T}(P^2=0) = \int_0^\infty\,dC\,\frac{1}{(1+C)^3}
                = \frac{1}{2}\,,
\label{hatThalf}
\end{equation}
so that, in the chiral limit, 
\begin{equation}
P_\rho\,{\cal T}_{\rho\mu\nu}(k_1,k_2) =
        \frac{\alpha_{\rm em}}{\pi}\,
\epsilon_{\mu\nu\rho\sigma} k_{1\rho} k_{2\sigma}\,.
\end{equation}

Hence the pseudoscalar piece of the neutral-pion Bethe-Salpeter amplitude
provides the only nonzero contribution to the divergence of the AVV
amplitude.  This contribution is just that identified with the ``triangle
anomaly'', and the result is {\it independent} of detailed information about
$\Gamma_\pi$ and $S(p)$.  It follows straightforwardly from
Eqs.~(\ref{hatTval}) and (\ref{hatThalf}) that
\begin{equation}
\Gamma_{\pi^0\to\gamma\gamma} = \frac{m_\pi^3}{16\pi} \frac{\alpha^2_{\rm
                em}}{\pi^2} \,{\cal T}(0)^2 
                = \frac{m_\pi^3}{64\pi} \left(\frac{\alpha_{\rm
                em}}{\pi \,f_\pi^0}\right)^2 \,.
\end{equation}

We emphasise that in obtaining the results in this section DCSB was crucial,
since it originates and is manifest in a nonzero value of $B_0$, in the
identity between $B_0$ and $E_\pi$, and in the other identities:
Eqs.~(\ref{awti})-(\ref{gwti}).

Our derivation is a generalisation of that in Ref.~\cite{cdrpion} and, to
make it simple, particular care was taken in choosing the momentum routing in
Eq.~(\ref{Trho}).  This was necessary because it is impossible to
simultaneously preserve the vector and axial vector Ward-Takahashi identities
for triangle diagrams in field theories with axial currents that are bilinear
in fermion fields.  This choice of variables ensures the preservation of the
vector Ward-Takahashi identity, which is tied to electromagnetic current
conservation.  With another choice of variables, surface terms arise that
modify the value of ${\cal R}(P^2=0)$.  However, these are always eliminated
by subtraction in any regularisation of the theory that ensures
electromagnetic current conservation.\cite{jackiw}

% \section{Electromagnetic pion form factor}
% \vspace*{0.5\baselineskip}
% 
% \hspace*{-\parindent}{\bf Electromagnetic pion form
% factor.}\hspace*{\parindent}
%
\section{Electromagnetic pion form factor}
As another example of the importance of $\Gamma_\pi$'s pseudovector
components, we consider the electromagnetic pion form factor, calculated in
the renormalised impulse approximation:
\begin{eqnarray}
\label{pipiA}
\lefteqn{(p_1 + p_2)_\mu\,F_\pi(q^2):= \Lambda_\mu(p_1,p_2)  }\\
& & \nonumber
= \frac{2 N_c}{N_\pi^2}\,\int\,\frac{d^4k}{(2\pi)^4}
        \,{\rm tr}_D\left[ \bar{\cal G}_\pi(k;-p_2)
S(k_{++})\, i\Gamma^\gamma_\mu(k_{++},k_{+-})\,S(k_{+-})\,
{\cal G}_\pi(k-q/2;p_1)\,S(k_{--})\right],
\end{eqnarray}
$k_{\alpha\beta}:= k + \alpha p_1/2 + \beta q/2$ and $p_2:= p_1 + q$.  Again,
no renormalisation constants appear explicitly in Eq.~(\ref{pipiA}) because
the renormalised dressed-quark-photon vertex, $\Gamma^\gamma_\mu$, satisfies
the vector Ward-Takahashi identity, Eq.~(\ref{vwti}).  This also ensures
current conservation:
\begin{equation}
(p_1-p_2)_\mu\,\Lambda_\mu(p_1,p_2)=0\,.
\end{equation}
We note that from the normalisation condition for ${\cal G}_\pi$,
Eq.~(\ref{pinorm}), and Eqs.~(\ref{vwti}) and (\ref{pipiA})
\begin{equation}
F(q^2=0)=1
\end{equation}
if, and only if, one employs a truncation in which $K$ is independent of $P$.
One such scheme is the rainbow-ladder truncation of Ref.~\cite{mr97}.

\subsection{Quark propagator}
To calculate $F_\pi(q^2)$ we employ an algebraic parametrisation of the
renormalised dressed-quark propagator that efficiently characterises many
essential and robust elements of the solutions obtained in studies of the
quark DSE.  This defines Eq.~(\ref{pipiA}) directly $\forall\, p_1^2,p_2^2$;
in particular at the pion mass shell.\footnote{The procedure actually
employed in Ref.~\cite{burkardt} can, at best, only reproduce our results.}
We introduce the dimensionless functions: $\bar\sigma_S(x):=
\lambda\,\sigma_S(p^2)$, $\bar\sigma_V(x):= \lambda^2\,\sigma_V(p^2)$, where
$p^2 = \lambda^2\,x$, $\lambda$ is a mass-scale, with
\begin{eqnarray}
\label{SSM}
\bar\sigma_S(\xi)  & =  & 
        2 \bar m {\cal F}(2 (\xi + \bar m^2))
        + {\cal F}(b_1\, \xi) {\cal F}(b_3\, \xi) 
        \left( b_0 + b_2 {\cal F}(\varepsilon\, \xi)\right)\,,\\
\label{SVM}
\bar\sigma_V(\xi) & = & \frac{2 (\xi+\bar m^2) -1 
                + e^{-2 (\xi+\bar m^2)}}{2 (\xi+\bar m^2)^2}
\end{eqnarray}
and ${\cal F}(y) := [1- \exp(-y)]/y$.  This five-parameter algebraic form,
where $\bar m$ is the $u/d$ current-quark mass, combines the effects of
confinement\footnote{The representation of $S(p)$ as an entire function is
motivated by the algebraic solutions of Eq.~(\protect\ref{gendse}) in
Refs.~\cite{munczekburden}.  The concomitant absence of a Lehmann
representation is a sufficient condition for
confinement.~\protect\cite{brs96,krein}} and DCSB with free-particle
behaviour at large, spacelike $p^2$.\footnote{At large-$p^2$: $\sigma_V(p^2)
\sim 1/p^2$ and $\sigma_S(p^2)\sim m/p^2$.  The parametrisation therefore
does not incorporate the additional $\ln p^2$-suppression characteristic of
QCD.  It is a useful but not necessary simplification, which introduces model
artefacts that are easily identified and accounted for.  $\varepsilon=0.01$
is introduced only to decouple the large- and intermediate-$p^2$ domains.}

The chiral limit vacuum quark condensate in QCD is~\cite{mrt98,mr97}:
\begin{eqnarray}
-\langle \bar q q \rangle^0_\mu & := &
\lim_{M^2\to\infty}\,Z_4(\mu^2,M^2)\,
        \frac{3}{4\pi^2}\,
        \int_0^{M^2}\,ds\,s\,\sigma_S^0(s)\,,
\label{condensate}
\end{eqnarray}
where at one-loop order
$ Z_4(\mu^2,M^2) = [\alpha(M^2)/\alpha(\mu^2)]^{\gamma_m ( 1 + \xi/3)}$,
with $\xi$ the covariant-gauge fixing parameter ($\xi=0$ specifies Landau
gauge) and $\gamma_m= 12/(33-2 N_f)$ the gauge-independent anomalous mass
dimension.  The $\xi$-dependence of $ Z_4(\mu^2,M^2) $ is just that required
to ensure that $\langle \bar q q \rangle^0_\mu$ is gauge independent.  The
parametrisation of Eq.~(\ref{SSM}) is a model that corresponds to the
replacement $\gamma_m \to 1$ in Landau gauge, in which case
Eq.~(\ref{condensate}) yields
\begin{eqnarray}
-\langle \bar q q \rangle^0_\mu & = & \lambda^3\,\ln\frac{\mu^2}{\Lambda_{\rm
QCD}^2}\,\frac{3}{4\pi^2}\, 
        \frac{b_0}{b_1\,b_3}\,.
\end{eqnarray}
This is the signature of DCSB in the model parametrisation and we calculate
the pion mass from
\begin{equation}
\label{gmor}
m_\pi^2\, f_\pi^2 = 2\,m \,\langle \bar q q \rangle^0_{1\, {\rm GeV}}\,.
\end{equation}
When all the components of $\Gamma_\pi$ are retained, Eq.~(\ref{gmor}) yields
an approximation to the pion mass found in a solution of the Bethe-Salpeter
equation that is accurate to 2\%~\cite{mr97}.

The model parameters are fixed by requiring a good description of a range of
pion observables.  This procedure explores our hypothesis that the bulk of
pion observables can be understood as the result of nonperturbative dressing
of the quark and gluon propagators.

\subsection{Pion Bethe-Salpeter Amplitude}
The Chebyshev moments of the scalar functions in $\Gamma_\pi(k;P)$ are, for
example,
\begin{equation}
E_\pi^i(k^2;P^2):= 
\frac{2}{\pi}\,\int_{0}^\pi\,d\beta\,\sin^2\beta\,U_i(\cos\beta)\,E_\pi(k;P)
\end{equation}
with $k\cdot P:= \sqrt{k^2 P^2}\,\cos\beta$, where $U_i(z)$ is a Chebyshev
polynomial of the second kind.  At large-$k^2$, independent of assumptions
about the form of $K$, one has~\cite{mr97}
\begin{equation}
\label{uvbeh}
E_\pi^0(k^2;P^2) \propto \,-\langle \bar q q \rangle^0_{k^2}
                \frac{\alpha(k^2)}{k^2}\,. 
\end{equation}
$F_\pi^0(k^2;P^2)$, $k^2\,G_\pi^0(k^2;P^2)$ and $k^2\,H_\pi^0(k^2;P^2)$ have
precisely the same behaviour; i.e., the asymptotic momentum-dependence of all
these functions is identical to that of $B_0(k^2)$.  This makes manifest the
``hard-gluon'' contribution to $F_\pi(q^2)$ in Eq.~(\ref{pipiA}).  Further,
in an asymptotically free theory, where a well constructed rainbow-ladder
truncation yields model-independent results at large-$k^2$~\cite{mr97},
\begin{equation}
k^2\,G_\pi^0(k^2;P^2) = 2 F_\pi^0(k^2;P^2)\,,\;
        k^2
\mathrel{\rlap{\lower4pt\hbox{\hskip0pt$\sim$}}
\raise2pt\hbox{$>$}} M_{\rm UV}^2\,,
\end{equation}
with $M_{\rm UV} := 10\, \Lambda_{\rm QCD}$.
% \footnote{We have not established whether this result persists for general
% forms of $K$.} 

In a model exemplar used in Ref.~\cite{mr97} the zeroth Chebyshev moments
provided results for $m_\pi$ and $f_\pi$ that were indistinguishable from
those obtained with the full solution.  Also $H_\pi \simeq 0$ and hence it
was quantitatively unimportant in the calculation of $m_\pi$ and $f_\pi$.  We
expect that these results are not specific to that particular model.  In the
latter case because the right-hand-side of Eq.~(\ref{gwti}) is zero and hence
in general there is no ``seed'' term for $H_\pi$.

These observations, combined with Eqs.~(\ref{bwti})--(\ref{gwti}), motivate a
model for $\Gamma_\pi$:
\begin{equation}
E_\pi(k;P) = \frac{1}{N_\pi} B_0(k^2)
\end{equation}
with 
$F_\pi(k;P)= E_\pi(k;P)/(110\,f_\pi) $,
$G_\pi(k;P)= 2 F_\pi(k;p)/[k^2 + M_{\rm UV}^2]$
and $H_\pi(k;P)\equiv 0$.
The relative magnitude of these functions at large $k^2$ is chosen to
reproduce the numerical results of Ref.~\cite{mr97}.

\subsection{Results}
We determined the model parameters by optimising a least-squares fit to
$f_\pi$, $r_\pi$ and $\langle \bar q q \rangle^0_{1\,{\rm GeV}}$, and a
selection of pion form factor data on the domain $q^2\in [0,4]\,$GeV$^2$.
The procedure does not yield a unique parameter set with, for example, the
two sets:
\begin{equation}
\label{params}
\begin{array}{lcccccc}
        & \lambda ({\rm GeV}) & \bar m & b_0 & b_1 & b_2 & b_3 \\
{\sf A} & 0.473               & 0.0127 & 0.329 & 1.51 & 0.429 & 0.430\,,\\
{\sf B} & 0.484               & 0.0125 & 0.314 & 1.63 & 0.445 & 0.405\,,
\end{array}
\end{equation}
providing equally good fits, as illustrated in
Table~\ref{tabres}.\footnote{The quark propagator obtained with these
parameter values is pointwise little different to that obtained in
Ref.~\protect\cite{cdrpion}.  One gauge of this is the value of the Euclidean
constituent quark mass; i.e., the solution of $p^2=M(p^2)^2$.  Here
$M^E_{u/d}= 0.32\,$GeV whereas $M^E_{u/d}= 0.30\,$GeV in
Ref.~\protect\cite{cdrpion}.  It is also qualitatively similar to the
numerical solution obtained in Ref.~\protect\cite{mr97}, where $M^E_{u/d}=
0.56\,$GeV.  Indeed, our results are not sensitive to details of the fitting
function: fitting with different confining, algebraic forms yields $S(p)$
that is pointwise little changed, and the same results for observables.}
There is a domain of parameter sets that satisfy our fitting criterion and
they are distinguished only by the calculated magnitude of the pion form
factor at large-$q^2$.  The two sets in Eq.~(\ref{params}) delimit reasonable
boundaries and illustrate the model dependence in our result.  With all
parameter sets in the acceptable domain, Eq.~(\ref{fpinpi}) is satisfied
exactly in the chiral limit, in which case we obtain $f_\pi^0= 0.090\,$GeV,
while at the fitted value of $m$, $N_\pi/f_\pi= 0.97$.

In our calculation $f_\pi r_\pi$ is 20\% too small.  This discrepancy cannot
be reduced in impulse approximation because the nonanalytic contributions to
the dressed-quark-photon vertex associated with $\pi$-$\pi$ rescattering and
the tail of the $\rho$-meson resonance are ignored~\cite{ewff}.  It can only
be eliminated if these contributions are included.  We have thus identified a
constraint on realistic, impulse approximation calculations: they should not
reproduce the experimental value of $f_\pi r_\pi$ to better than $\sim20$\%,
otherwise the model employed has unphysical degrees-of-freedom.

\begin{figure}
\centering{\
\epsfig{figure=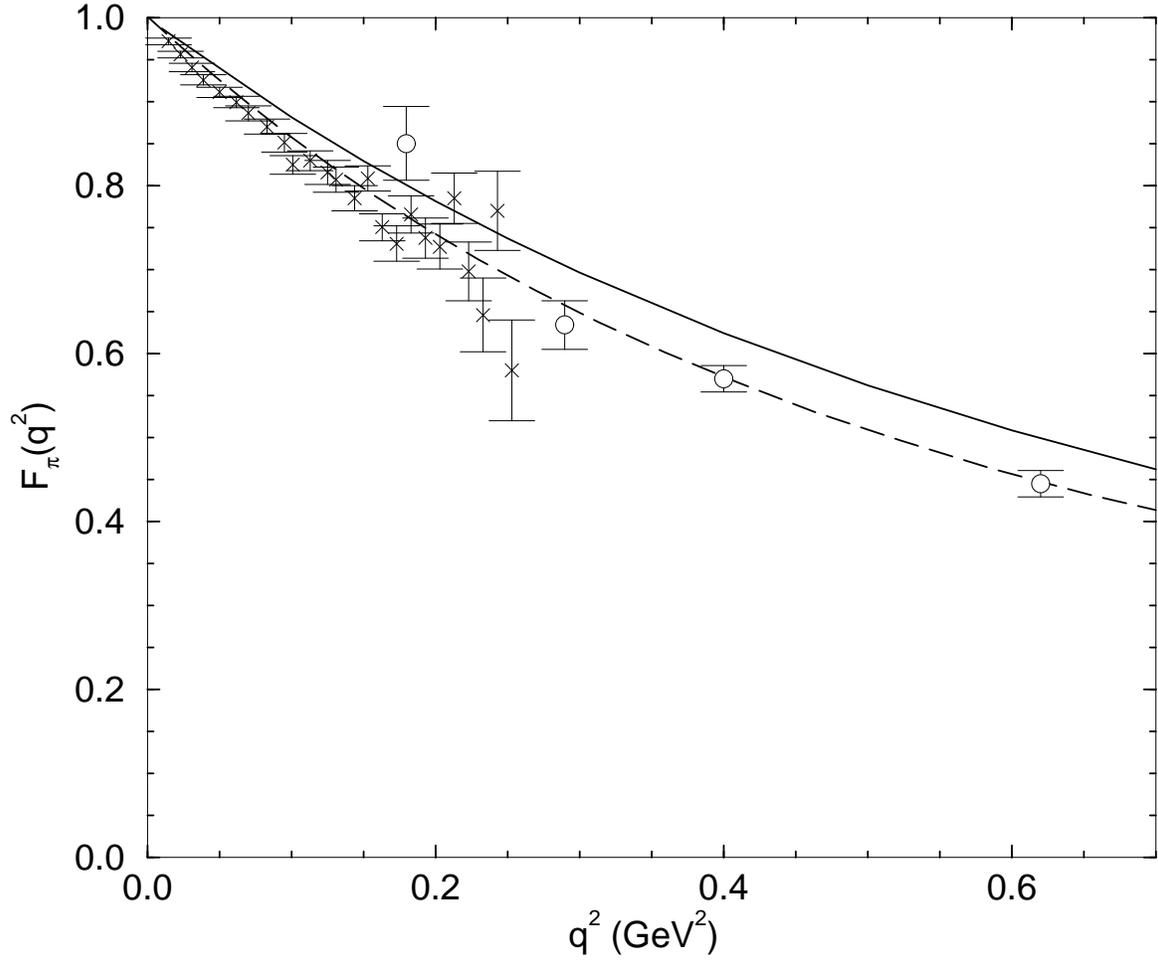,height=13.0cm}}
\caption{Calculated pion form factor compared with data at small $q^2$.  The
data are from Refs.~\protect\cite{amend} (crosses) and \protect\cite{bebek}
(circles).  The solid line is the result obtained when the pseudovector
components of the pion Bethe-Salpeter amplitude are included, the dashed-line
when they are neglected~\protect\cite{cdrpion}.  On the scale of this figure,
both parameter sets in Eq.~(\protect\ref{params}) yield the same calculated
result. 
\label{smallF}}
\end{figure}

\begin{figure}
\centering{\
\epsfig{figure=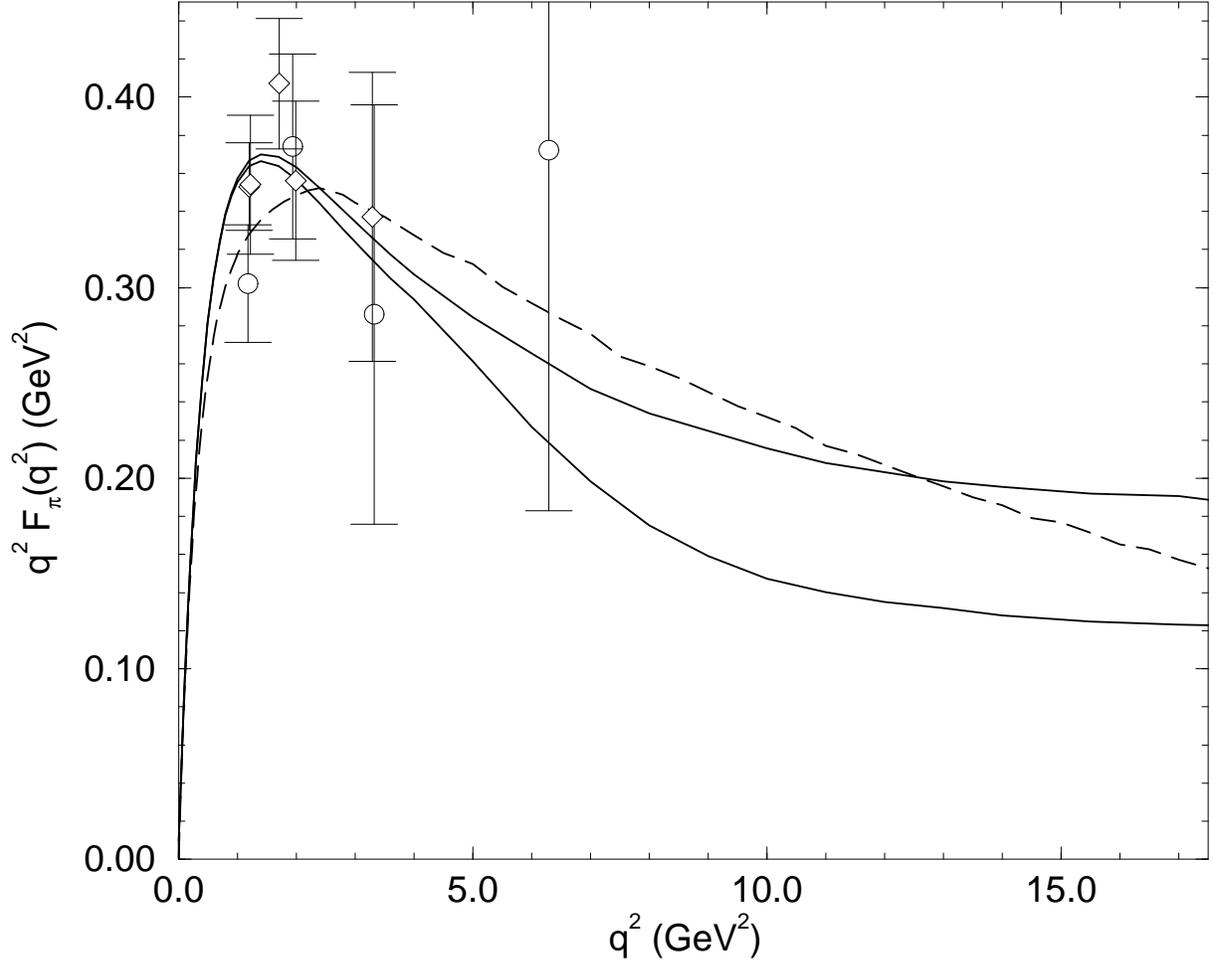,height=13.0cm}}
\caption{Calculated pion form factor compared with the largest $q^2$ data
available: diamonds - Ref.~\protect\cite{bebek}; and circles -
Ref.~\protect\cite{bebekB}.  The solid lines are the results obtained when
the pseudovector components of the pion Bethe-Salpeter amplitude are included
(lower line - set {\sf A} in Eq.~(\protect\ref{params}); upper line - set
{\sf B}), the dashed-line when they are neglected~\protect\cite{cdrpion}.
\label{midF}}
\end{figure}

Our calculated pion form factor is compared with available data in
Figs.~\ref{smallF} and \ref{midF}.  It is also compared with the result
obtained in Ref.~\cite{cdrpion}, wherein the calculation is identical {\em
except} that the pseudovector components of the pion were neglected.
% \footnote{Our results are not sensitive to details of the fitting function.
% Fitting with different confining, algebraic forms yields $S(p)$ that is
% pointwise little changed, and the same results for observables.} 
Figure~\ref{smallF} shows a small, systematic discrepancy between both
calculations and the data at low $q^2$, which is due to the underestimate of
$r_\pi$ in impulse approximation.\footnote{Just as in the present
calculation, $f_\pi r_\pi = 0.25$ in Ref.~\protect\cite{cdrpion}.  However,
the mass-scale is fixed so that $f_\pi=0.084$, which is why this result
appears to agree better with the data at small-$q^2$: $r_\pi$ is larger.}
The results obtained with or without the pseudovector components of the pion
Bethe-Salpeter amplitude are quantitatively the same, which indicates that
the pseudoscalar component, $E_\pi$, is dominant in this domain.

The increasing uncertainty in the experimental data at intermediate $q^2$ is
apparent in Fig.~\ref{midF}, as is the difference between the results
calculated with or without the pseudovector components of the pion
Bethe-Salpeter amplitude.  These components provide the dominant contribution
to $F_\pi(q^2)$ at large pion energy because of the multiplicative factors:
$\gamma\cdot P$ and $\gamma\cdot k\,k\cdot P$, which contribute an additional
power of $q^2$ in the numerator of those terms involving $F^2$, $FG$ and
$G^2$ relative to those proportional to $E$.  Using the method of
Ref.~\cite{cdrpion} and the model-independent asymptotic behaviour indicated
by Eq.~(\ref{uvbeh}) we find
\begin{equation}
\label{FUV}
F_\pi(q^2) \propto \frac{\alpha(q^2)}{q^2}\,
        \frac{(-\langle \bar q q\rangle^0_{q^2})^2}{f_\pi^4}\,;
\end{equation}
i.e., $q^2 F_\pi(q^2) \approx {\rm const.}$, up to calculable $\ln
q^2$-corrections.  If the pseudovector components of $\Gamma_\pi$ are
neglected, the additional numerator factor of $q^2$ is missing and one
obtains~\cite{cdrpion} $q^4 F_\pi(q^2)\approx {\rm const.}$

In our model the behaviour identified in Eq.~(\ref{FUV}) becomes apparent at
$q^2 \mathrel{\rlap{\lower4pt\hbox{\hskip0pt$\sim$}}
\raise2pt\hbox{$>$}} 2\,M_{UV}^2
$.
This is the domain on which the asymptotic behaviour indicated by
Eq.~(\ref{uvbeh}) is manifest.  Our calculated results, obtained with the two
sets of parameters in Eq.~(\ref{params}), illustrate the model dependent
uncertainty:
\begin{equation}
\left.q^2 F_\pi(q^2)\right|_{q^2 \sim 10-15\,{\rm GeV}^2} \sim 0.12 -
0.19\,{\rm GeV}^2\,.
\end{equation}
This uncertainty arises primarily because the model allows for a change in
one parameter to be compensated by a change in another.  In our example:
$b_2^{\sf B} > b_2^{\sf A}$ but $b_0^{\sf B}+b_2^{\sf B}=b_0^{\sf A}+b_2^{\sf
A}$; and $b_1^{\sf A}\,b_3^{\sf A}= b_1^{\sf B}\,b_3^{\sf B}$.  This allows
an equally good fit to low-energy properties but alters the
intermediate-$q^2$ behaviour of $F_\pi(q^2)$.  In a solution of
Eq.~(\ref{gendse}) these coefficients of the $1/p^4$ and $1/p^6$ terms are
correlated and such compensations cannot occur.

As a comparison, evaluating the leading-order perturbative-QCD result with
the asymptotic quark distribution amplitude:
$\phi_{\rm as}(x) := \surd 3\,f_\pi\, x (1-x)$, 
yields 
$q^2 F_\pi(q^2) = 16\,\pi f_\pi^2 \,\alpha(q^2) \approx 0.15\,{\rm GeV}^2$, 
assuming a value of $\alpha(q^2\sim 10\,{\rm GeV}^2)\approx 0.3$.  However,
the perturbative analysis neglects the anomalous dimension accompanying
condensate formation.\footnote{For example, Eqs.~(\ref{bwti})-(\ref{gwti})
are not satisfied in Ref.~\protect\cite{FJ}.}

% \section{Conclusions}
% \vspace*{0.5\baselineskip}
% 
% \hspace*{-\parindent}{\bf Conclusions.}\hspace*{\parindent}
%
\section{conclusions}
Using the Dyson-Schwinger equations it is straightforward to show that, as a
consequence of the dynamical chiral symmetry breaking mechanism, the pion is
a nearly-massless, pseudoscalar, quark-antiquark bound
state~\cite{mrt98,mr97}.  As a corollary, the complete pion Bethe-Salpeter
amplitude necessarily contains pseudovector and pseudotensor components,
which are always qualitatively important.  In model studies, the quantitative
effect of these components can be obscured in the calculation of many pion
observables; i.e., within a judiciously constructed framework, applied at
low- to intermediate-energy, their effect can be absorbed into the values of
the model parameters~\cite{pctrev,cdrpion}.  However, they are crucial to a
proper realisation of anomalous current divergences, crucial to obtaining a
uniformly accurate connection between the low- and high-energy domains, and
they provide the dominant contribution to the electromagnetic pion form
factor at $q^2 > 10\,$GeV$^2$.

% \acknowledgements 
%
% \vspace*{0.5\baselineskip}
% 
% \hspace*{-\parindent}{\bf Acknowledgments.}\hspace*{\parindent}
%
\acknowledgements
This work was supported by the US Department of Energy, Nuclear Physics
Division, under contract number W-31-109-ENG-38 and benefited from the
resources of the National Energy Research Scientific Computing Center.

%______________________________ References ______________________________

%....................
\begin{table}[h]
\begin{tabular}{cll}
        &       Calculated      &       Experiment \\\hline
$f_\pi$ &       0.092$\,$GeV    &       0.092  \\
$(-\langle \bar q q\rangle^0_{1\,{\rm GeV}})^{1/3}$
        &       0.236           &       0.236 $\pm$ 0.008~\cite{derek} \\
$m_{u/d}$ &     0.006           &       0.008 $\pm$ 0.004~\cite{pdg96} \\
$m_\pi$ &       0.1387          &       0.1385 \\
$r_\pi$ &       0.55$\,$fm      &       0.663 $\pm$ 0.006~\cite{amend}\\
$r_\pi f_\pi$ & 0.25$\,$(dimensionless)&       0.310 $\pm$ 0.003
\end{tabular}
\caption{A comparison between our calculated values of low-energy pion
observables and experiment or, in the case of $(-\langle \bar q
q\rangle^0_{1\,{\rm GeV}})^{1/3}$ and $m_{u/d}$, the values estimated using
other theoretical tools.  Each of the parameter sets in
Eq.~(\protect\ref{params}) yields the same calculated values.  For
consistency with Ref.~\protect\cite{mr97}, we use $\Lambda_{\rm
QCD}=0.234\,$GeV throughout.
\label{tabres}}
\end{table}
%--------------------Figures
%--------------------
%____________________________________________________________________________
\end{document}